\documentclass[preprint,showpacs,aps]{revtex4} 
\input{epsf}  
\usepackage{graphicx}
\usepackage{epic}
\usepackage{graphpap}
\usepackage{epic}
\usepackage{eepic}
\usepackage{color}
\usepackage{amssymb}
\usepackage{amsmath}
 
\begin{document}  

\title{
Stochastic Analysis of Dimerization Systems
} 
\author{Baruch Barzel and Ofer Biham}  
\affiliation{  
Racah Institute of Physics,   
The Hebrew University,   
Jerusalem 91904,   
Israel}  

\newcommand{\NA}
{
\langle N_A \rangle
}

\newcommand{\NB}
{
\langle N_B \rangle
}

\newcommand{\ND}
{
\langle N_D \rangle
}

\newcommand{\NAss}
{
\langle N_A \rangle^{\rm ss}
}

\newcommand{\NBss}
{
\langle N_B \rangle^{\rm ss}
}

\newcommand{\NDss}
{
\langle N_D \rangle^{\rm ss}
}

\newcommand{\NANB}
{
\langle N_A N_B \rangle
}

\newcommand{\NAsNB}
{
\langle {N_A}^2 N_B \rangle
}

\newcommand{\NANBs}
{
\langle N_A {N_B}^2 \rangle
}

\newcommand{\NAs}
{
\langle {N_A}^2 \rangle
}

\newcommand{\NAc}
{
\langle {N_A}^3 \rangle
}

\newcommand{\R}
{
\langle R \rangle
}

\newcommand{\Rss}
{
\langle R \rangle^{\rm ss}
}

\newcommand{\arrow}[1]
{
\overset{#1}{\longrightarrow}
}

\newcommand{\aeff}
{
a_{\rm eff}
}

\newcommand{\Reff}
{
\R_{\rm eff}
}

\newcommand{\gammaeff}
{
\gamma_{\rm eff}
}

\newcommand{\sigmass}
{
\sigma
}

\begin{abstract}  

The process of dimerization, 
in which two monomers bind to each other and form a dimer,
is common in nature.
This process can be modeled using rate equations,
from which
the average copy numbers of the reacting monomers 
and of the product dimers can then be obtained.
However, the rate equations apply only when 
these copy numbers are large.
In the limit of small copy numbers 
the system becomes dominated by fluctuations,
which are not accounted for by the rate equations.
In this limit one must use stochastic methods such as 
direct integration of the master equation
or Monte Carlo simulations.
These methods are computationally intensive and rarely 
succumb to analytical solutions. 
Here we use the recently introduced moment equations 
which provide a highly simplified stochastic 
treatment of the dimerization
process.
Using this approach, we 
obtain an analytical solution for the copy numbers
and reaction rates both under steady state 
conditions and in the time-dependent case.
We analyze three different dimerization processes:
dimerization without dissociation,
dimerization with dissociation
and
hetero-dimer formation. 
To validate the results we compare them with 
the results obtained from the master equation
in the stochastic limit
and with those obtained from the rate equations
in the deterministic limit.
Potential applications of the results 
in different physical contexts
are discussed.
  
\end{abstract}  
  
 
\maketitle

\section{Introduction}
\label{Introduction}

Dimerization is a common process in 
physical, chemical and biological systems.
In this process, two identical units (monomers)
bind to each other and form a dimer
($A + A \rightarrow A_2$).
This is a special case of a more general
reaction process (hetero-dimerization) of the form
$A + B \rightarrow AB$. 
Dimerization may appear
either as an isolated process
or 
incorporated in
a more
complex reaction network.
The modeling of dimerization systems is commonly done using
rate equations, which incorporate the mean-field approximation.
These equations describe the time evolution 
of the concentrations of the monomers and the dimers.
Assuming that the system is spatially homogeneous,
these concentrations can be expressed either in terms
of the copy numbers per unit volume or in terms of the
total copy number of each molecular species in the system.
The rate equations are reliable when the 
copy numbers of the reacting
monomers in the system are sufficiently large for the 
mean-field approximation to apply.
However, when the copy numbers of the reacting monomers 
are low, the system becomes highly fluctuative, 
and the rate equations are no longer suitable.
Therefore the analysis of dimerization processes 
under conditions of low copy numbers
requires the use of stochastic methods
\cite{vanKampen1981,Gardiner2004}.
These methods include the direct integration of
the master equation 
\cite{Biham2001,Green2001}, 
and 
Monte Carlo (MC) simulations 
\cite{Gillespie1976,Gillespie1977,Tielens1982,Newman1999,Charnley2001}.
The master equation
consists of coupled differential equations 
for the probabilities of all the possible microscopic states
of the system.
These equations are typically solved by 
numerical integration.
However,
in some simple cases, the steady state probabilities
can be obtained by analytical methods
\cite{Green2001,Biham2002}. 
The difficulty with the master equation is that it consists
of a large number of equations,
particularly if the dimerization process is a part of 
a larger network.
This severely limits its usability for the analysis of complex
reaction networks
\cite{Stantcheva2002,Stantcheva2003}.
Monte Carlo simulations provide a stochastic implementation
of the master equation, following the actual temporal
evolution of a single instance of the system.
The mean copy numbers of the reactants and the reaction
rates are obtained by averaging over an ensemble of such instances.
In both methods, there is no closed form expression for the
time dependence of the copy numbers and reaction rates.

Recently, a new method for the stochastic modeling of reaction networks
was developed, which is
based on moment equations 
\cite{Lipshtat2003,Barzel2007a,Barzel2007}. 
The moment equations 
are much more efficient than the master equation.
They consist of only one equation for each reactive species, 
one equation for each reaction rate, 
and in certain cases 
one equation for each product species.
Thus, the number of moment equations 
that describe a given chemical network
is 
comparable to the number of rate equations,
which consist of one equation for each species.
Moreover, 
unlike the rate equations, 
the moment equations 
are linear equations.
In some cases, this feature enables to 
obtain an analytical solution for the time dependent 
concentrations. 

In this paper,
we apply the moment equations to the analysis
of dimerization systems with fluctuations.
These equations are accurate even in the limit of
low copy numbers, where fluctuations are large
and the rate equations fail.
We show how to obtain an analytical solution for 
the time dependent concentrations of the reactant
and product species
as well as for the reaction rate.
We identify and characterize the
different dynamical regimes of the system
as a function
of the parameters.
We examine the validity of our solution 
by comparison to the results obtained from the master equation.
The analysis is performed for three
variants of the system: dimerization, dimerization
with dissociation and hetero-dimer formation.

%
%

The paper is organized as follows. 
In Sec. \ref{sec2} we analyze the dimerization process
using the moment equations and provide an analytical solution
for the time-dependent concentrations.
In Sec. \ref{sec3} we extend the analysis to the
case in which dimers may dissociate. 
In Sec. \ref{sec4} we generalize the analysis to 
the formation of hetero-dimers.
The results are summarized
and potential applications are discussed 
in Sec. \ref{sec5}.

\section{Dimerization systems}
\label{sec2}

Consider a system of molecules, denoted by
$A$, which diffuse and react on 
a surface or in a liquid solution.
Molecules are produced or added to the system
at a rate 
$g$ (s$^{-1}$),
and degrade at a rate  
$d_1$ (s$^{-1}$).
When two molecules encounter each other
they may bind and form the dimer 
$D$. 
The rate constant for the dimerization process is denoted by
$a$ (s$^{-1}$).
For simplicity, we assume that 
the product molecule,
$D$,
is non-reactive and undergoes degradation at a rate 
$d_2$ (s$^{-1}$). 
The chemical processes in this 
system can be described by

\begin{eqnarray}
&& \varnothing \arrow{g} A    \nonumber \\
&& A \arrow{d_1} \varnothing  \nonumber \\
&& A + A \arrow{a} D          \nonumber \\
&& D \arrow{d_2} \varnothing.
\label{eq:processes1}
\end{eqnarray}

\subsection{Rate Equations}

The dimerization system described above is characterized 
by the average copy number of the monomers,
$\NA$, 
and by the average copy number of the dimers,
$\ND$. 
Denoting
the average dimerization rate 
by
$\R$ (s$^{-1}$),
the rate equations for this system take the form

\begin{eqnarray}
\frac{d\NA}{dt} &=& g - d_1 \NA - 2\R  \nonumber \\
\frac{d\ND}{dt} &=& - d_2 \ND + \R.
\label{eq:rate1}    
\end{eqnarray}

\noindent
These equations include a term for each process 
which appears in
Eq. (\ref{eq:processes1}).
The factor of 
$2$
in the reaction term of the first equation
accounts for the fact that the dimerization process removes
two $A$ molecules, 
producing one dimer.
The dimerization rate,
$\R$,
is proportional to 
the number of pairs of $A$ molecules
in the system,
$\NA(\NA-1)/2$,
where the factor of $1/2$ is absorbed into the
rate constant $a$.
As long as the copy number of $A$ molecules is large,
it can be approximated by

\begin{equation}
\R = a{\NA}^2.
\label{eq:Rrate1}
\end{equation}
   
\noindent
Eqs. (\ref{eq:rate1}) form a closed set 
of two non-linear differential equations.
Their steady state solution 
is

\begin{eqnarray}
\NAss &=& 
\frac{d_1}{4a}
\left(
-1 + \sqrt{1 + 8 \gamma}
\right)
\nonumber \\
\NDss &=& 
\frac{d_1^2}{16 a d_2}
\left(
-1 + \sqrt{1 + 8 \gamma}
\right)^2,
\label{eq:ss_rate1}
\end{eqnarray}

\noindent
where 
$\gamma$
is the reaction strength parameter given by

\begin{equation}
\gamma = \frac{ag}{d_1^2}.
\label{eq:gamma}
\end{equation}

\noindent
Two limits can be identified. 
In the limit where
$\gamma \gg 1$, 
the steady state dimerization rate satisfies
$\Rss \simeq g/2$, 
and the steady state dimer population is
$\NDss \simeq g/2d_2$.
This means
that almost all the monomers that are generated
end up in dimers and 
the monomer degradation process becomes irrelevant.
Therefore,
this limit is referred to 
as the reaction-dominated regime.
The degradation-dominated limit is obtained when
$\gamma \ll 1$.
In this limit 
$\NAss \simeq g/d_1$,
$\Rss \simeq a g^2 / {d_1}^2 = \gamma g$
and 
$\NDss \simeq a g^2/({d_1}^2 d_2)$,
namely most of the monomers that are generated 
undergo degradation and only a small fraction
end up in dimers.

The time-dependent solution for the population size of the 
$A$
molecules can be obtained from the first equation in
Eqs. (\ref{eq:rate1}).  
The result is


\begin{equation}
\NA = \NAss - 
\frac{1}{2 a \tau}
\left(
1 + C e^{t/\tau}
\right)^{-1},
\label{eq:rate_solution1}
\end{equation}

\noindent
where
$\tau = 1/\sqrt{d_1^2 + 8ag}$
is the relaxation time 
and the parameter
$C$
is determined by the initial conditions.
In the reaction-dominated regime, where
$a g \gg d_1^2$, 
the relaxation time converges to
$\tau \simeq 1/\sqrt{8ag}$.
In the degradation-dominated regime, 
it approaches
$\tau \simeq 1/d_1$.

The rate equation analysis is valid as long as 
the copy numbers of the reactive
molecules are sufficiently large
\cite{Lederhendler2008}.
In the limit in which the copy number 
$\NA$ 
of the monomers 
is reduced to order unity or less, 
the rate equations 
[Eqs. (\ref{eq:rate1})] 
become unsuitable.
This limit can be reached in two situations:
when the monomer concentration is very low or
when the volume of the system is very small.
In the limit of low copy number of the monomers,
the system becomes dominated by fluctuations
which are not accounted for by the rate equations.
A useful characterization of the system is
given by the system-size parameter

\begin{equation}
N_0 = \frac{g}{d_1},
\label{eq:N0}
\end{equation}

\noindent
which approximates the copy number
of the monomers in case that
the dimerization is suppressed.
The parameter 
$N_0$
provides an upper limit for the monomer population size
under steady state conditions.
It can be used to characterize the dynamical regime 
of the system.
In the limit where 
$N_0 \gg 1$
the copy number of the monomers is typically 
large and the rate equations are reliable.
However, when 
$N_0 \lesssim 1$
the system may become dominated by fluctuations. 
In this regime, the rate equations
fail to account for the population sizes and the dimerization rate.
In Fig. \ref{fig1} we present a schematic illustration of
the parameter space in terms of $\gamma$ and $N_0$,
identifying the four dynamical regimes. 

\subsection{Moment Equations}

To obtain a more complete description of the dimerization process,
which takes the fluctuations into account, 
we present the master equation approach.
The dynamical variables
of the master equation are the probabilities
$P(N_A,N_D)$
of having a population of 
$N_A$ 
monomers 
and 
$N_D$
dimers 
in the system. 
The master equation 
for the dimerization system
takes the form

\begin{eqnarray}
\label{eq:master1}
\frac{dP(N_A,N_D)}{dt} &=&
g [P(N_A - 1,N_D) - P(N_A,N_D)]
\nonumber \\ &+&
d_1 [(N_A + 1)P(N_A + 1,N_D) - N_A P(N_A,N_D)]
\\ 
&+&
d_2 [(N_D + 1)P(N_A,N_D + 1) - N_D P(N_A,N_D)]
\nonumber \\ 
&+&
a [(N_A + 2)(N_A + 1)P(N_A + 2,N_D - 1) - N_A(N_A - 1)P(N_A,N_D)].
\nonumber
\end{eqnarray}

\begin{figure}
\includegraphics[width=12cm]{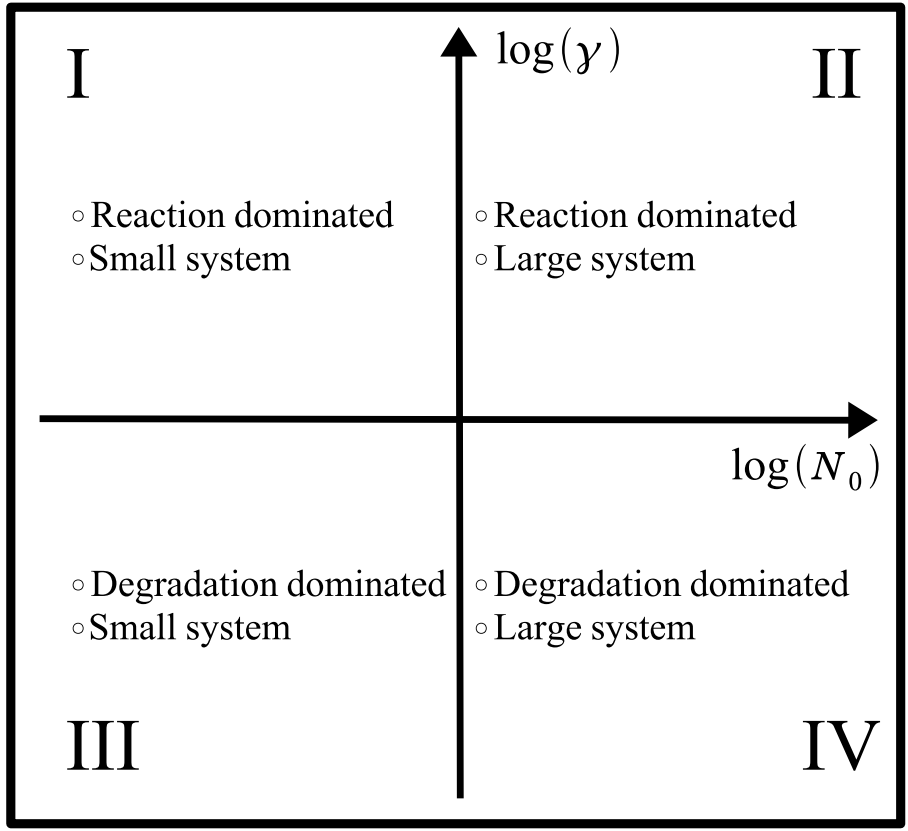}
\caption{
Dimerization systems can be classified into four different regimes,
characterized by the system size parameter,
$N_0 = g/d_1$,
and by the reaction strength parameter
$\gamma = ag/d_1^2$.
In the large system limit, where
$N_0 \gg 1$ (quadrants II and IV),
the monomer copy number is typically large,
and the fluctuation level of the system is low.
In the small system limit, where 
$N_0 \ll 1$ (quadrants I and III),
the monomer copy number is low, and the system becomes
dominated by fluctuations.
The reaction strength parameter characterizes the dominant
dynamical process in the system.
In the limit where
$\gamma \gg 1$ (quadrants I and II)
the process of dimerization is dominant,
and the degradation is suppressed.
In the limit where
$\gamma \ll 1$ (quadrants III and IV)
the degradation process is dominant, and only a small
fraction of the monomers undergo dimerization.
}
\label{fig1}
\end{figure}

\begin{figure}
\includegraphics[width=12cm]{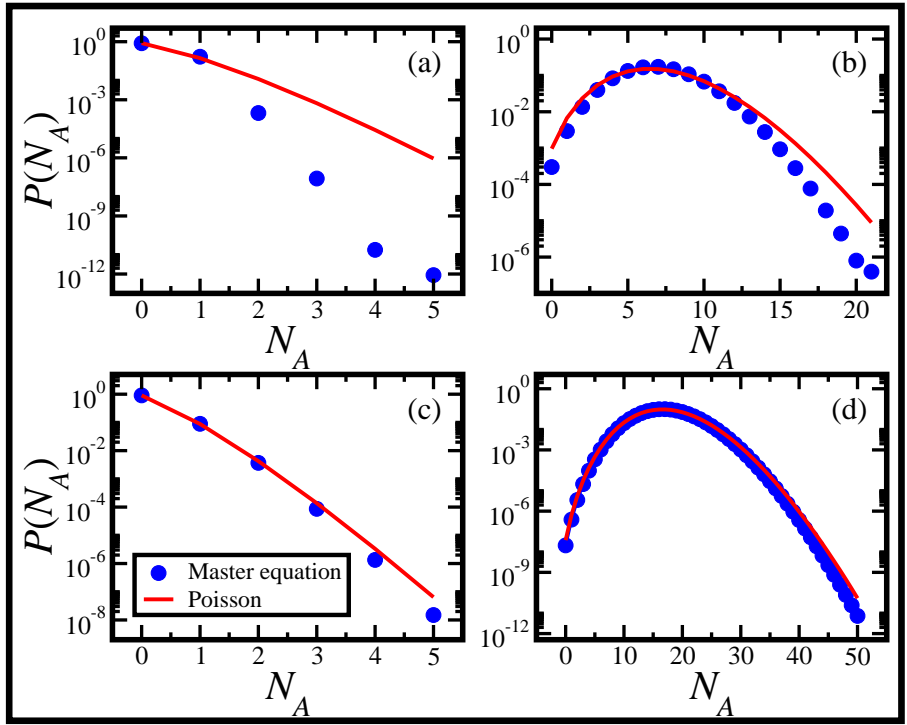}
\caption{
(color online)
The steady state probability distribution
$P(N_A)$, 
as obtained from the master equation (circles)
for
(a) small degradation-dominated system;
(b) small reaction-dominated system; 
(c) large degradation-dominated system
and
(d) large reaction-dominated system. 
For the degradation-dominated systems [(c) and (d)],
$P(N_A)$
can be fitted by
a Poisson distribution (solid lines).
For the reaction-dominated systems [(a) and (b)], 
$P(N_A)$
significantly deviates from the Poisson distribution.
}
\label{fig2}
\end{figure}

\noindent
The first term on the right hand side 
accounts for the addition or formation of 
$A$
molecules. 
The second and third terms account 
for the degradation of 
$A$
and 
$D$
molecules,
respectively.
The last term describes the reaction process, in which two
$A$ 
molecules are annihilated and one 
$D$ molecule is formed.
The dimerization rate is 
proportional to the number of pairs of
$A$
molecules in the system, given by 
$N_A (N_A - 1) / 2$.
Therefore, the dimerization rate can be expressed in terms
of the moments of 
$P(N_A,N_D)$
as 

\begin{equation}
\R = a
\left(
\NAs - \NA
\right),
\label{eq:Rmaster1} 
\end{equation}

\noindent
where the moments are defined by

\begin{equation}
\langle N_A^n N_D^m \rangle = 
\sum_{\substack{N_A = 0\\N_D = 0}}^{\infty}
{N_A^n N_D^m P(N_A, N_D)},
\label{eq:moments}
\end{equation}

\noindent 
and $n$ and $m$ are integers.
Note that in the stochastic formulation, the 
expression used for the dimerization rate,
$\R$,
is different than in the deterministic approach 
[Eq. (\ref{eq:Rrate1})].
The two expressions are equal in case that
$P(N_A)$ is a Poisson distribution, for
which the mean and the variance are equal.
The master equation (\ref{eq:master1}) 
for the dimerization system can be analytically solved 
to obtain the steady state probabilities
$P(N_A)$.
This solution can be found at refs. 
\cite{Green2001,Biham2002}.
However, an
analytical solution for the time dependent case is currently 
not available.
For dimerization systems in the degradation-dominated limit, the
probability distribution
$P(N_A)$
approaches the Poisson distribution.
However as 
$\gamma$
increases, and the system enters the reaction-dominated regime,
$P(N_A)$
becomes different from Poisson.
In Fig. \ref{fig2} 
we present the marginal probability distributions 
$P(N_A)$ (circles)
as obtained from the master equation for four choices of the
parameters, each in one of
the four regimes shown in Fig. \ref{fig1}.
In the limit of 
$N_0 \ll 1$
and 
$\gamma \gg 1$ (a)
the system is in the reaction dominated regime
(quadrant I in Fig. \ref{fig1}), 
and
the results obtained from the master equation deviate from  
Poisson (solid lines).
Here the parameters are
$g = 0.5$,
$d_1 = 2$,
$a = 200$
and
$d_2 = 10$ (s$^{-1}$).
The distribution shown in (b),
where
$N_0 \gg 1$
and
$\gamma \gg 1$
(quadrant II in Fig. \ref{fig1}), 
also deviates from the Poisson distribution. 
Here the parameters are
$g = 100$,
$d_1 = 1$,
$a = 1$
and
$d_2 = 10$ (s$^{-1}$).
In the limit of 
$N_0 \ll 1$
and 
$\gamma \ll 1$ (c)
the system is in the degradation-dominated regime
(quadrant III in Fig. \ref{fig1}), 
and correspondingly
$P(N_A)$
coincides with the 
Poisson distribution.
Here the parameters are
$g = 0.5$,
$d_1 = 5$,
$a = 1$
and
$d_2 = 10$ (s$^{-1}$).
Finally, in (d)  
$N_0 \gg 1$
and
$\gamma \ll 1$ 
(quadrant IV in Fig. \ref{fig1}), 
the system is dominated by degradation and
as before,
the master equation results coincide with the Poisson distribution.
Here the parameters are
$g = 10$,
$d_1 = 1$,
$a = 5 \times 10^{-3}$
and
$d_2 = 10$ (s$^{-1}$).

In addition to the analytical solution mentioned above, 
the master equation 
[Eq. (\ref{eq:master1})]
can also be integrated numerically using standard 
steppers such as 
the Runge Kutta method
\cite{Acton1970,Press1992}.
In numerical simulations, 
one has to truncate the master equation 
in order to keep the number of equations finite. 
This is achieved by setting upper cutoffs 
$N_A^{\rm max}$ 
and 
$N_D^{\rm max}$ 
on the numbers of $A$ and $D$ molecules,
respectively.
This truncation is valid if the probability for the number of molecules of 
each type to exceed the cutoff is vanishingly small.

The population sizes of the 
$A$ 
and
$D$ 
molecules and the dimerization rate are 
expressed in terms of the first moments 
of 
$P(N_A, N_D)$
and one of its second moments,
$\NAs$.
Therefore, a closed set of equations for the time derivatives of
these first and second moments could directly provide all the information
needed in order to evaluate the population sizes and the dimerization rate
\cite{Lipshtat2003}.
Such equations can be obtained from the master equation using
the identity

\begin{equation}
\frac{d\langle N_A^n N_D^m \rangle}{dt} = 
\sum_{\substack{N_A = 0\\N_D = 0}}^{\infty}
{N_A^n N_D^m \dot P(N_A, N_D)}.
\label{eq:moments_dot}
\end{equation}

\noindent 
Inserting 
the time-derivative
$\dot P(N_A, N_D)$ 
according to Eq. (\ref{eq:master1}),
one obtains the moment equations. 
The equations for the average copy numbers are

\begin{eqnarray}
\frac{d\NA}{dt} & = & g - d_1\NA - 2\R
\nonumber \\
\frac{d\ND}{dt} & = & -d_2\ND + \R,
\label{eq:moment1_no_R}
\end{eqnarray}

\noindent
while the equation for
the dimerization rate is

\begin{equation}
\frac{d\R}{dt} = (2ag+4a^2) \NA + (10a - 2d_1)\R - 4a^2 \NAc.
\label{eq:Rmoment1_not_truncated}
\end{equation}

\noindent
Eqs. (\ref{eq:moment1_no_R}) 
have the same form as the
rate equations (\ref{eq:rate1}). 
However the term for the dimerization rate,
$\R$,
as appears in the moment equations is different from the analogous
term in the rate equations 
[Eq. (\ref{eq:Rrate1})].

Eqs. (\ref{eq:moment1_no_R}), 
together with Eq. (\ref{eq:Rmoment1_not_truncated}),
are a set of coupled differential equations,
which are linear in terms of the moments.
Although we have written the equations only for the 
relevant first and second moments, 
the right hand side of 
Eq. (\ref{eq:Rmoment1_not_truncated}) 
includes the third moment for which we have no equation.
In order to close the set of equations one must express this
third moment in terms of the first and second moments.
Different expressions have been proposed.
For example, in the context of birth-death processes
the relation 
$\NAc = \NAs \NA$
was used 
\cite{McQuarrie1967}.
This choice makes the moment equations 
nonlinear, 
which might affect their 
stability.
Another common choice is to
assume that the third central moment is 
zero (which is exact for symmetric distributions)
and use this relation to express
the third moment in terms of the 
first and second moments
\cite{Gomez-Uribe2007}.
Here we use a different approach.
We set up the closure condition
by imposing a highly restrictive cutoff 
on the master equation.
The cutoff is set at $N_{A}^{\rm max} = 2$.
This is the minimal cutoff 
that still enables the dimerization process to take place.
Under this cutoff, 
one obtains the following relation
between the first three moments
\cite{Lipshtat2003} 

\begin{equation}
\NAc = 3\NAs -2 \NA.
\label{eq:third_moment}
\end{equation}

\noindent
Using this result, one can bring the moment equations
[Eqs. (\ref{eq:moment1_no_R}) - (\ref{eq:Rmoment1_not_truncated})]
into a closed form:

\begin{eqnarray}
\frac{d\NA}{dt} &=& g - d_1\NA - 2\R
\nonumber \\
\frac{d\ND}{dt} &=& -d_2\ND + \R
\nonumber \\
\frac{d\R}{dt}  &=& 2ag\NA - 2(d_1 + a)\R. 
\label{eq:moment1}
\end{eqnarray}

\noindent
Numerical integration of these equations provides all the required moments,
from which the population sizes and the dimerization rate are obtained.

\subsubsection{Steady State Analysis}

The steady-state solution of the moment 
equations 
takes the form

\begin{eqnarray}
\NAss &=& 
\frac{g(a + d_1)}{2ag + d_1 a + d_1^2} 
\nonumber \\
\NDss &=& 
\frac{ag^2}{d_2(2ag + d_1 a + d_1^2)}
\nonumber \\
\Rss &=& 
\frac{ag^2}{2ag + d_1 a + d_1^2}.
\label{eq:ss_moment1}
\end{eqnarray}  

\noindent
In the limit of very small copy numbers 
the approximation appearing in 
Eq. (\ref{eq:third_moment}) 
is valid. 
Thus, in this limit the
moment equations 
provide accurate results, both for the population sizes (first moments)
and for the dimerization rate (involving a second moment).
To evaluate the validity of the moment equations in the limit of 
large copy numbers, we compare 
Eqs. (\ref{eq:ss_moment1}) 
with the solution of the rate equations 
(\ref{eq:ss_rate1}), 
which is valid in this limit.
Consider the large-system limit, where
$N_0 \gg 1$
[Eq. (\ref{eq:N0})].
In this limit, 
the common term in
the denominators in 
Eqs. (\ref{eq:ss_moment1}) approaches 
${d_1}^2 (2 \gamma + 1)$,
where 
$\gamma$ 
is the reaction strength parameter, 
given by Eq. (\ref{eq:gamma}).
Thus, in the degradation-dominated limit, 
where
$\gamma \ll 1$,
the steady state solution of the moment equations approaches

\begin{eqnarray}
\NAss &=& 
\frac{g}{d_1} 
\nonumber \\
\NDss &=& 
a \frac{g^2}{d_2 {d_1}^2}
\nonumber \\
\Rss &=& 
a \frac{g^2}{{d_1}^2}.
\label{eq:ss_moment1_gamma_small}
\end{eqnarray}

\noindent
Here we use the fact that in order to 
satisfy both the large-system limit
($N_0 \gg 1$),
and the degradation-dominated limit
($\gamma \ll 1$),
one must also require
$d_1 \gg a$.
The results appearing in 
Eqs. (\ref{eq:ss_moment1_gamma_small}) are consistent
with the results of the rate equations in this limit.
We conclude that the moment equations are also reliable for large 
populations under the condition that the 
system is in the degradation-dominated regime.
To test the results of the moment equations 
for large systems in 
the reaction-dominated regime
we examine the case of
$\gamma \gg 1$.
Here Eqs. (\ref{eq:ss_moment1}) are reduced to

\begin{eqnarray}
\NAss &=& 
\frac{a + d_1}{2a} 
\nonumber \\
\NDss &=& 
\frac{g}{2 d_2}
\nonumber \\
\Rss &=& 
\frac{g}{2}.
\label{eq:ss_moment1_gamma_large}
\end{eqnarray}

\noindent
In this limit,
the monomer copy number
$\NAss$,
obtained from the moment equations, 
does not match the 
rate equation result.
Nevertheless, the results for the dimer population size,
$\NDss$, 
and for the dimerization rate 
$\Rss$,
do converge to the results obtained from the rate equations.

We thus conclude that the accuracy 
of the moment equations is maintained 
well beyond the small system limit.
The equations provide accurate results for the dimer copy number,
$\NDss$,  
and for the dimerization rate,
$\Rss$,
for both small and large systems.
As for the monomer copy number, 
the moment equations provide an accurate 
description in all limits, 
except for the limit where both
$N_0 \gg 1$
and
$\gamma \gg 1$
(quadrant II in Fig. \ref{fig1}).
In Table \ref{tab1}
we present a characterization of the different
dynamical regimes and the applicability of the
moment equations for the evaluation of
the copy numbers and the dimerization rate
in each regime.

\begin{figure}
\includegraphics[width=12cm]{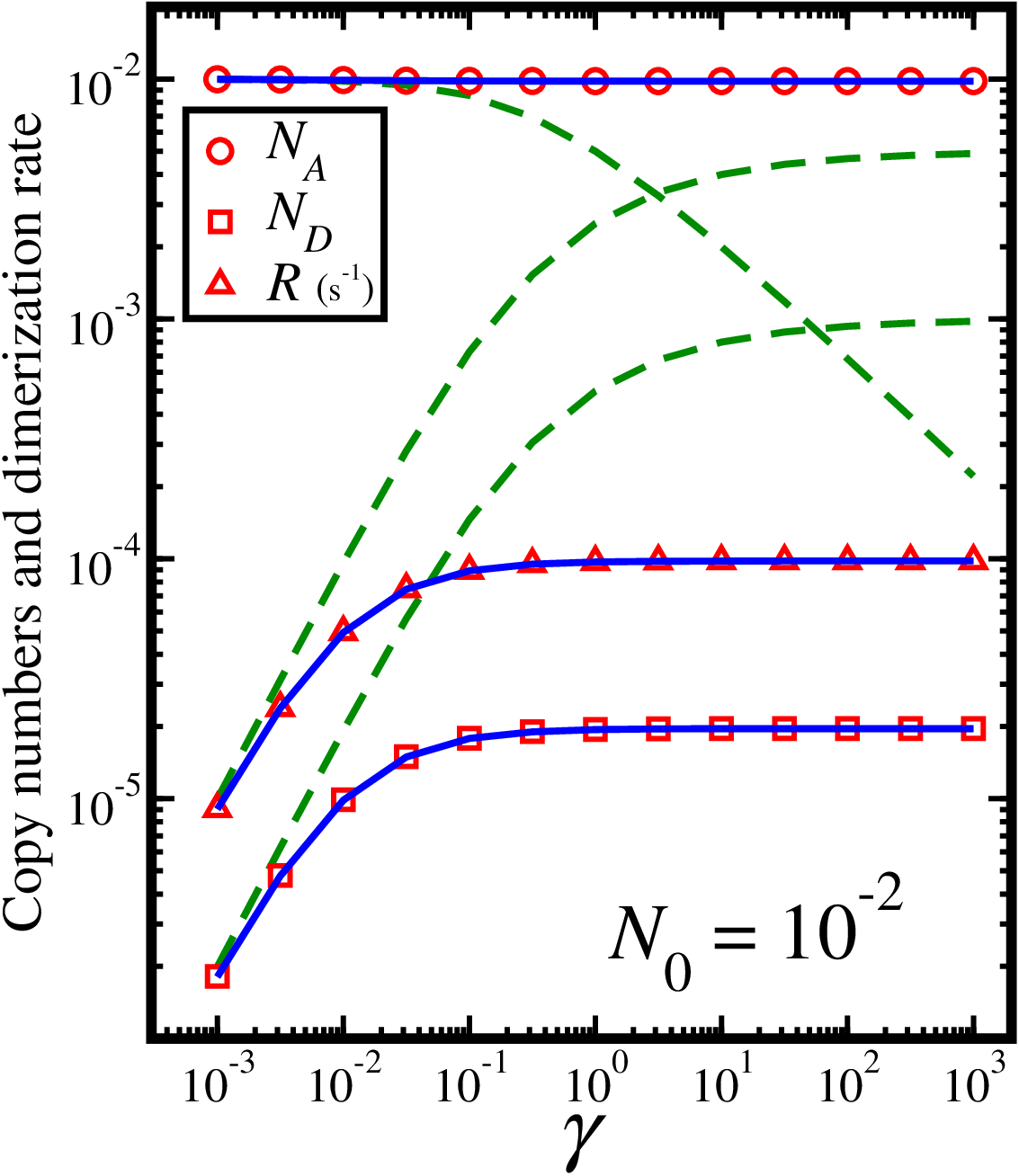}
\caption{
(color online)
The average monomer copy number,
$\NAss$ (circles),
the average dimer copy number,
$\NDss$ (squares)
and the dimerization rate,
$\Rss$ (triangles),
versus the reaction strength parameter,
$\gamma$,
as obtained from the moment equations
under steady state conditions.
The parameters used here satisfy the small
system limit,
namely
$N_0 = 10^{-2}$.
The moment equation results are in perfect agreement 
with the results obtained
from the master equation (solid lines).
However, the rate equation results (dashed lines)
show significant deviations.
This is because in the limit of small 
monomer copy number the system is dominated by
fluctuations, which are not accounted 
for by the rate equations.
Note that  
$\NAss$ 
and
$\NDss$ 
are dimensionless while   
$\Rss$
is units of 
s$^{-1}$.
}
\label{fig3}
\end{figure}

In Fig. \ref{fig3} 
we present 
the monomer copy number
$\NAss$ (circles),
the dimer copy number
$\NDss$ (squares)
and the dimerization rate 
$\Rss$ (triangles),
as obtained from the moment equations, 
versus
the reaction strength parameter,
$\gamma$.
The rate constants are
$g = 0.01$,
$d_1 = 1$
and
$d_2 = 5$
(s$^{-1}$).
The reaction rate, 
$a$, 
is varied.
These parameters satisfy the small-system limit
$N_0 \ll 1$.
The moment equation results are in excellent agreement 
with those obtained from the master equation (solid lines).
However, since the populations are small, 
the results of the rate equations
show deviations (dashed lines).
In Fig. \ref{fig4} we present 
$\NAss$ (circles),
$\NDss$ (squares)
and
$\Rss$ (triangles),
as obtained from the moment equations, 
versus
the reaction strength parameter,
$\gamma$.
Here the rate constants are
$g = 10^{3}$,
$d_1 = 0.1$
and
$d_2 = 0.1$
(s$^{-1}$).
As before,
the reaction rate, 
$a$, 
is varied.
These parameters satisfy the large-system limit
$N_0 \gg 1$,
and thus the rate equation results (dashed lines) are accurate.
Although the populations are large for the entire parameter range
displayed,
the results of the moment equations are in excellent agreement 
with those obtained from the rate equations.
The only deviation appears in the results for the monomer population
in the limit  
$\gamma \gg 1$.
For the parameters used in this simulation it was impractical to simulate 
the master equation.

In any chemical reaction it is important to 
characterize the extent to which
fluctuations are significant. 
From the 
master equation, one can evaluate the fluctuation level in the 
monomer copy number, 
given by the variance 

\begin{equation}
\sigma^2 = \NAs - \NA^2,
\label{eq:sigma}
\end{equation}

\noindent
where $\sigma$ is the standard deviation.
The problem is that the expression for 
$\sigma$ includes
the first moment
$\NA$,
which
is not always accurately accounted for by the moment equations.
However, when the copy number is sufficiently large, 
the rate equations 
apply, and thus one can extract the value of 
$\NA^2$
in this limit from the rate equations. 
On the other hand, the moment equations account correctly for
the second moment
$\NAs$
by
$\NAs = \R/a + \NA$.
Using this relation, 
the result at steady state is

\begin{eqnarray}
\sigmass^2 &=&
\left\{
\begin{array}{ccccr}
\frac
{g
\left[
d_1^3 + a^2(d_1+g) +a (2d_1^2 + d_1 g + 2 g^2)
\right]}
{\left(
2 a g + a d_1 + d_1^2
\right)^2}
& & {\rm for} & & N_0 \le 1
\\ \\ 
\frac
{g (g + d_1 + a)}
{2 a g + a d_1 + d_1^2} - 
\frac{d_1^2}{16 a}
\left(
-1 + \sqrt{1 + 8 \gamma}
\right)^2
& & {\rm for} & & N_0 > 1.
\end{array}
\right.
\label{eq:moment_rate_sigma}
\end{eqnarray}

\noindent
In Fig. \ref{fig5} we present
the coefficient of variation
$\sigmass / \NAss$ (circles), 
as obtained from 
Eq. (\ref{eq:moment_rate_sigma})
versus the system size parameter
$N_0$. 
The parameters are
$d_1=1$,
$a=1$,
$d_2=5$ (s$^{-1}$)
and $g$ 
is varied.
Here 
$\NAss$ 
was extracted from the moment equations for
$N_0 \le 1$,
and from the rate equations for
$N_0 > 1$.
In the small-system limit 
($N_0 \ll 1$),
the average fluctuation becomes much larger than 
$\NAss$.
The system is thus dominated by fluctuations.
As the system size increases, 
$\sigmass$
becomes small with respect to
$\NAss$,
implying that the system enters the deterministic regime.
In order to validate our results,
we compare them with results obtained
from the master equation (solid line).
For 
$N_0 < 1$
the agreement is perfect, as in this limit the moment equations are 
expected to be accurate.
A slight deviation appears for 
$N_0 > 1$,
where 
$\sigmass$
is constructed by 
combining results obtained from
the moment 
equations and from the rate equations.
In both limits,
Eq. (\ref{eq:moment_rate_sigma})
is found to provide a good approximation 
for the fluctuation level
of the system. 

\subsubsection{Time-Dependent Solution}

The time dependent solution for 
$\NA$
can be obtained by solving the two coupled equations for
$\NA$ 
and for
$\R$
in Eqs. (\ref{eq:moment1}).
The equation for
$\ND$
receives input from these two equations.
However, the dimers are the final products of this
network and 
$\ND$
has no effect on
$\NA$ 
and
$\R$.
Thus the equations for $\NA$ and $\R$ can be decoupled
from the equation for $\ND$.
One obtains a set of two 
coupled linear differential equations 
of the form

\begin{equation}
\dot{ \vec{ N}} = {\bf M} \vec N + \vec b,
\label{eq:moment1_matrixform}
\end{equation}

\noindent
where
$\vec N = (\NA, \R)$,
$\vec b = (g, 0)$
and the matrix
${\bf M}$ 
is

\begin{eqnarray}
{\bf M} = 
\left(
\begin{array}{cc}
- d_1    & -2
\\
2 a g    & -2(d_1 + a)
\end{array}
\right).
\label{eq:M}
\end{eqnarray}

\noindent
The two eigenvectors of the matrix 
${\bf M}$
are given by

\begin{eqnarray}
\vec v_1 = 
\left(
\begin{array}{c}
\frac
{2a + d_1 - \omega}
{4 a g}
\\
1
\end{array}
\right); 
&&
\vec v_2 = 
\left(
\begin{array}{c}
\frac
{2a + d_1 + \omega}
{4 a g}
\\
1
\end{array}
\right),
\label{eq:eigenvectors}
\end{eqnarray}

\noindent
where 
$\omega = \sqrt{4 a^2 + d_1^2 + 4 a d_1 - 16 a g}$.
The corresponding eigenvalues are

\begin{eqnarray}
-\frac{1}{\tau_1} = 
\frac{1}{2}
(-2 a - 3 d_1 - \omega);
&&
-\frac{1}{\tau_2} = 
\frac{1}{2}
(-2 a - 3 d_1 + \omega).
\label{eq:eigenvalues}
\end{eqnarray}

\noindent
Using 
the matrix
${\bf Q} = (\vec v_1, \vec v_2)$,
one can write 
Eq. (\ref{eq:moment1_matrixform})
as

\begin{equation}
{\bf Q}^{-1} \dot{ \vec{ N}} =
{\bf Q}^{-1} {\bf M} {\bf Q} 
{\bf Q}^{-1} \vec N + 
{\bf Q}^{-1} \vec b.
\label{eq:moment1_matrixform_Qued}
\end{equation}
 
\noindent
The result is a set of two un-coupled differential 
equations of the form

\begin{eqnarray}
\dot{ \vec{ f}} &=&
\left(
\begin{array}{cc}
- \frac{1}{\tau_1}  & 0
\\
0  & - \frac{1}{\tau_2}
\end{array}
\right)
\vec f + \vec k, 
\label{eq:moment1_uncoupled}
\end{eqnarray}

\noindent
where 
$\vec f = {\bf Q}^{-1} \vec N$
and
$\vec k = {\bf Q}^{-1} \vec b$.
The solution of Eq. (\ref{eq:moment1_uncoupled}) is 

\begin{eqnarray}
\vec f(t) &=&
\left(
\begin{array}{c}
k_1 \tau_1 + C_1 e^{-{t}/{\tau_1}}
\\
k_2 \tau_1 + C_2 e^{-{t}/{\tau_2}}
\end{array}
\right),
\label{eq:moment1_uncoupled_solution}
\end{eqnarray}
  
\noindent
where 
$C_1$ and $C_2$
are arbitrary constants.
Multiplying 
Eq. (\ref{eq:moment1_uncoupled_solution})
from the left hand side by the matrix 
${\bf Q}$
one obtains the time dependent solution of 
Eq. (\ref{eq:moment1_matrixform}),
which is

\begin{eqnarray}
\NA &=&
\frac{g(a + d_1)}
{2 a g + a d_1 + d_1^2}
+
{\bf Q}_{1,1} C_1 e^{-{t}/{\tau_1}} +
{\bf Q}_{1,2} C_2 e^{-{t}/{\tau_2}}
\nonumber \\
\R &=&
\frac{a g^2}
{2 a g + a d_1 + d_1^2}
+
{\bf Q}_{2,1} C_1 e^{-{t}/{\tau_1}} +
{\bf Q}_{2,2} C_2 e^{-{t}/{\tau_2}}.
\label{eq:moment1_time_solution}
\end{eqnarray}

\noindent
The first terms on the right hand side of 
Eqs. (\ref{eq:moment1_time_solution})
are the steady state solutions 
$\NAss$
and 
$\Rss$
as they appear in Eqs. (\ref{eq:ss_moment1}).
The second and third terms represent the time-dependent parts of 
$\NA$
and 
$\R$.
These terms exhibit 
an exponential decay with two 
characteristic relaxation times,
$\tau_1$
and
$\tau_2$. 
Practically, since 
$\tau_1 < \tau_2$,
the effective relaxation time for the monomers is 
$\tau_A = \tau_2$.

In the limit of small copy numbers, where
$N_0 \ll 1$,
Eqs. (\ref{eq:moment1_time_solution}) 
account correctly for the copy numbers and for
the reaction rates.
In this limit 
(where $g \ll d_1$),
one obtains 
$\tau_A \simeq 1/d_1$.
In the limit of large copy numbers, where
$N_0 \gg 1$,
one has to distinguish between degradation-dominated and
reaction-dominated systems.
Consider a degradation-dominated system, where
$N_0 \gg 1$
and
$\gamma \ll 1$.
These two conditions require that
$d_1 \gg a$.
As before,
the effective relaxation time is approximated by
$\tau_A \simeq 1/d_1$.
This result is consistent with the 
results of the rate equations in this regime
[Eq. (\ref{eq:rate_solution1})].
A peculiar result arises in the reaction-dominated 
regime, in the limit of
large populations.
In this limit
$\omega \simeq \sqrt{4 a^2 - 16 a g}$.
For 
$a < 4 g$
the parameter
$\omega$
becomes a purely imaginary number.
This result leads to spurious oscillations in the solution 
presented in
Eqs. (\ref{eq:moment1_time_solution}).
As shown for the steady state solution 
[Eq. (\ref{eq:ss_moment1})],
the moment equations consistently fail in the limit of reaction-dominated
systems with large copy numbers.
To obtain the relaxation time in this limit, one can rely on the 
results obtained from the rate equations, which give
$\tau_A \simeq 1/\sqrt{8 a g}$.
The relaxation times in all the different limits are summarized in 
Table \ref{tab2}. 

Finally, we refer to the time evolution of the dimer population
$\ND$.
The equation for  
$\ND$
is the second equation in 
Eqs. (\ref{eq:moment1}), where 
$\R$ 
is to be taken from 
Eqs. (\ref{eq:moment1_time_solution}).
The solution of this equation takes the form

\begin{equation}
\ND = \NDss + 
\tilde C_1 e^{-{t}/{\tau_1}} +
\tilde C_2 e^{-{t}/{\tau_2}} +
C_3 e^{-{t}/{\tau_3}},
\label{eq:ND_time_solution}
\end{equation} 

\noindent
where
$\NDss$
is taken from
Eqs. (\ref{eq:ss_moment1}),
$\tilde C_i = {\bf Q}_{2,i} C_i / [d_2 - (1/\tau_i)]$,
$\tau_3 = 1/d_2$ 
and
$C_3$ 
is an arbitrary constant.
The effective relaxation time for the copy number
of the dimer product depends on the value of
$\tau_3$.
If 
$\tau_3 < \tau_A$,
the copy number of the dimers relaxes rapidly.
Thus, 
the time required for
the dimers to reach steady state is determined by 
the monomer relaxation time, namely
$\tau_D \simeq \tau_A$.
In the opposite case,
where
$\tau_3 > \tau_A$,
the monomer population reaches steady state quickly, 
and the production rate of 
the dimers acts in effect as a constant generation rate.
Correspondingly, the relaxation time for 
$\ND$
in this limit is approximated by
$\tau_D \simeq 1/d_2$
(Table \ref{tab2}).

In Fig. \ref{fig6}(a) 
we present the time evolution of
$\NA$ (circles)
$\ND$ (squares)
and 
$\R$ (triangles),
as obtained from the moment equations.
The parameters are
$g = 2 \times 10^{-3}$,
$d_1 = 0.05$,
$a = 100$
and
$d_2 = 5$
(s$^{-1}$).
These parameters correspond to the small system limit and to the
reaction-dominated regime (quadrant I in Fig. \ref{fig1}).
The moment equations (symbols) are in perfect agreement with the 
master equation (solid lines).
The rate equations (dashed lines) deviate from the stochastic results 
both in evaluating the steady state values of 
$\NA$, $\ND$ and $\R$,
and in predicting the relaxation times of 
$\NA$ and $\R$.
According to the rate equations, 
this relaxation time should be
$\tau_A \simeq 1/\sqrt{8 a g} \simeq 0.8$ (s),
while according to the stochastic description
$\tau_A \simeq 1/d_1 \simeq 20$ (s).
In 
Fig. \ref{fig6}(b) 
we present results for a system in the large population 
limit and in the regime of reaction-dominated kinetics
(quadrant II in Fig. \ref{fig1}). 
Here the parameters are
$g = 10$,
$d_1 = 0.5$,
$a = 1$
and
$d_2 = 10$
(s$^{-1}$).
Under these conditions the moment equations fail to produce the correct 
time transient, and give rise to an oscillatory solution (symbols).
In this regime the results obtained from the rate equations
(dashed lines) are accurate 
and coincide with
the master equation
results (solid lines).
Note that even in this case the 
moment equations provide the correct values 
for the dimer production rate, 
$\R$,
and for the dimer population,
$\ND$,
under steady state conditions.
In Fig. \ref{fig6}(c) 
the parameters are 
$g = 0.01$,
$d_1 = 2$,
$a = 10$
and
$d_2 = 10$
(s$^{-1}$).
These parameters satisfy the small system 
limit and are in the kinetic regime
dominated by degradation
(quadrant III in Fig. \ref{fig1}).
The results are in perfect agreement with those obtained from the 
master equation (solid lines).
However, the rate equations (dashed lines), 
although displaying similar relaxation times,
show significant deviations in the steady state values of
$\ND$ and $\R$.
In 
Fig. \ref{fig6}(d) 
we present results for the case of a 
large system, where 
the parameters are
$g = 10$,
$d_1 = 0.5$,
$a = 10^{-3}$
and
$d_2 = 0.05$
(s$^{-1}$).
These parameters correspond to a system in the degradation-dominated regime
(quadrant IV in Fig. \ref{fig1}).
Although in this system the copy numbers are large, the
results obtained from the moment equations (symbols) 
are in perfect agreement with those obtained 
from the master equation (solid lines) and from the rate equations
(dashed lines).
Here the relaxation time for the monomer population is
$\tau_A \simeq 1/d_1$ 
and for the dimer population it is
$\tau_D \simeq 1/d_2$.

\section{Dimerization-Dissociation Systems}
\label{sec3}

To generalize the discussion of the previous Section we now consider 
the case where the dimer product
$D$
may undergo dissociation 
into two monomers,
at a rate 
$u$ (s$^{-1}$).
The chemical processes in this system are thus

\begin{eqnarray}
&& \varnothing \arrow{g} A    \nonumber \\
&& A \arrow{d_1} \varnothing  \nonumber \\
&& A + A \arrow{a} D          \nonumber \\
&& D \arrow{d_2} \varnothing  \nonumber \\
&& D \arrow{u} A + A.
\label{eq:processes2}
\end{eqnarray}

\subsection{Rate Equations}

The rate equations for this reaction take the form

\begin{eqnarray}
\frac{d\NA}{dt} &=& g - d_1 \NA - 2\R + 2 u \ND \nonumber \\
\frac{d\ND}{dt} &=& - (d_2 + u) \ND + \R,
\label{eq:rate2}    
\end{eqnarray}

\noindent
where $\R = a \NA^2$.
These equations are similar to 
Eqs. (\ref{eq:rate1}),
except for the 
$u$ terms
which account for the dissociation.
We define the effective reaction rate constant as
$\aeff = a[d_2/(u+d_2)]$
such that
the effective dimerization rate is
$\Reff = \aeff \NA^2$.
Under steady state conditions,
Eqs. (\ref{eq:rate2}) 
can be written as

\begin{eqnarray}
g - d_1 \NA - 2\Reff &=& 0 
\nonumber \\
\Reff - d_2 \ND &=& 0.
\label{eq:rate2_eff}
\end{eqnarray}

\noindent
They take
the same form as 
Eqs. (\ref{eq:rate1}), for dimerization 
without dissociation,
under steady state conditions.
The steady state solution for these equations is

\begin{eqnarray}
\NAss &=& 
\frac{d_1}{4 \aeff}
\left(
-1 + \sqrt{1 + 8 \gammaeff}
\right)
\nonumber \\
\NDss &=& 
\frac{d_1^2}{16 \aeff d_2}
\left(
-1 + \sqrt{1 + 8 \gammaeff}
\right)^2,
\label{eq:ss_rate2}
\end{eqnarray}

\noindent
where

\begin{equation}
\gammaeff = \frac{g \aeff}{d_1^2}
\label{eq:gammaeff}
\end{equation}

\noindent
is the effective reaction strength parameter.
In the limit where 
$d_2 \gg u$, 
most dimers undergo degradation. 
The dissociation process is suppressed, 
and the effective reaction rate constant is
$\aeff \simeq a$,
namely the solution approaches that of 
dimerization without dissociation.
In the limit where 
$d_2 \ll u$,
most of the produced dimers end 
up dissociating into monomers, 
and correspondingly
$\aeff \rightarrow 0$.
In this limit, the dimerization and dissociation processes
reach a balance. 
The effective dimerization rate vanishes
and $\NAss \simeq g/d_1$. 

\subsection{Moment Equations}

In order to conduct a stochastic analysis we present the master equation
for the dimerization-dissociation system, which takes the form

\begin{eqnarray}
\frac{dP(N_A,N_D)}{dt} &=&
g [P(N_A - 1,N_D) - P(N_A,N_D)]
\nonumber \\ &+&
d_1 [(N_A + 1)P(N_A + 1,N_D) - N_A P(N_A,N_D)]
\nonumber \\ &+&
d_2 [(N_D + 1)P(N_A,N_D + 1) - N_D P(N_A,N_D)]
\nonumber \\ &+&
a [(N_A + 2)(N_A + 1)P(N_A + 2,N_D - 1) - N_A(N_A - 1)P(N_A,N_D)]
\nonumber \\ &+&
u [(N_D + 1)P(N_A - 2,N_D + 1) - N_D P(N_A,N_D)].
\label{eq:master2}
\end{eqnarray}
 
\noindent
This equation resembles 
Eq. (\ref{eq:master1}), 
except for the last 
term which accounts 
for the dissociation process.
The master equation can be solved 
numerically by imposing suitable
cutoffs, 
$N_A^{\rm max}$
and
$N_D^{\rm max}$.
However an analytical solution is currently unavailable.
To obtain a much simpler stochastic 
description of this system we refer to the
moment equations.
Following the same steps as 
in the previous Section,
we impose the minimal cutoffs on the master 
equation, that enable all the required processes to
take place.
More specifically, we choose
$N_A^{\rm max} = 2$
in order to enable the dimerization.
We do not limit the copy number of the dimer, $N_D$.
However, we do not allow $N_A \ne 0$ and 
$N_D \ne 0$ simultaneously, because $A$ and $D$
molecules do not react with each other.
These cutoffs reproduce the closure condition of
Eq. (\ref{eq:third_moment}).
They also gives rise to another closure condition,
which is needed here, namely
$\langle N_A N_D \rangle = 0$.
The closed set of moment equations
takes the form

\begin{eqnarray}
\frac{d\NA}{dt} &=& g - d_1\NA - 2\R + 2u\ND
\nonumber \\
\frac{d\ND}{dt} &=& - (u + d_2)\ND + \R
\nonumber \\
\frac{d\R}{dt}  &=& 2ag\NA - 2(d_1 + a)\R + 2au\ND.
\label{eq:moment2}
\end{eqnarray}

\noindent
The steady state solution of these equations is

\begin{eqnarray}
\NA &=& 
\frac{g(\aeff + d_1)}{2g\aeff + d_1 \aeff + d_1^2} 
\nonumber \\
\ND &=& 
\frac{\aeff g^2}{d_2(2g\aeff + d_1 \aeff + d_1^2)}
\nonumber \\
\R &=& 
\frac{ag^2}{2g\aeff + d_1 \aeff + d_1^2}.
\label{eq:ss_moment2}
\end{eqnarray}  

\noindent
Note that this solution resembles the steady state solution shown in
Eqs. (\ref{eq:ss_moment1}), except for the replacement of 
$a$
by
$\aeff$.
As before, the validity of the moment 
equations can be characterized by the 
system size parameter,
$N_0$,
and by the effective reaction strength parameter
$\gammaeff$.
For small systems, where
$N_0 \ll 1$,
the approximation underlying the moment equations is valid,
and thus the moment equations provide accurate results for 
$\NA$, 
$\ND$
and
$\R$.
In the limit of large systems, where
$N_0 \gg 1$,
the validity of the moment equations can be evaluated by comparison with 
the rate equations. 
Two limits are observed.
In the degradation-dominated limit, where
$\gammaeff \ll 1$,
the solution obtained from the moment equations 
(\ref{eq:ss_moment2})
converges to the solution obtained from the rate equations
(\ref{eq:ss_rate2}).
The moment equations are thus valid in this limit for 
the monomer copy number,
$\NA$,
as well as for the dimer copy number,
$\ND$,
and its production rate,
$\R$.
However, for large systems in the reaction-dominated limit, where
$\gammaeff \gg 1$,
the moment equations converge to the rate equations only for 
$\ND$
and 
$\R$.
In this limit the monomer population size,
$\NA$, 
is not correctly accounted for by the moment equations.
In conclusion, 
the validity of the moment equations 
is the same as in the case of dimerization without dissociation
(Table \ref{tab1}) 
under the substitution
$\gamma \rightarrow \gammaeff$.

In Fig. \ref{fig7} 
we present
$\NAss$ (circles),
$\NDss$ (squares) 
and 
$\Rss$ (triangles),
as obtained from the moment equations for 
the dimerization-dissociation system
versus the effective reaction strength,
$\gammaeff$.
Here the parameters are
$g = 0.02$,
$d_1 = 1$,
$a = 2500$
and
$d_2 = 1$ 
(s$^{-1}$).
The variation of $\gammaeff$
along the horizontal axis was achieved by  
varying the
dissociation rate constant,
$u$.
For these parameters the system is in the 
small population limit, namely
$N_0 \ll 1$.
The moment equation results are found to be in 
perfect agreement with the results 
obtained from the master equations (solid lines).
However, the rate equations (dashed lines) show significant deviations
for a wide range of parameters.
These deviations are largest when the dimerization process is dominant 
($\gammaeff > 1$) 
as the effects of stochasticity become important. 
In Fig. \ref{fig8} 
we present
$\NAss$ (circles),
$\NDss$ (squares) 
and 
$\Rss$ (triangles),
as obtained from the moment equations, 
versus the effective reaction strength,
$\gammaeff$.
Here the parameters are
$g = 1000$,
$d_1 = 1$,
$a = 1$
and
$d_2 = 1$ 
(s$^{-1}$).
The dissociation rate constant,
$u$,
was varied.
For these parameters the system is 
in the large population limit, namely
$N_0 \gg 1$.
Although the population sizes of the 
monomer and of the dimer are large, the 
moment equations are in perfect 
agreement with the rate equations (dashed lines)
in the limit of
$\gammaeff \ll 1$.
For  
$\gammaeff \gg 1$
this agreement is maintained for the dimer population size 
and for its production rate.
In this limit the monomer population 
size is not accounted for by the moment equations.
Slight deviations in $\ND$ and $\R$
appear within a narrow range around
$\gammaeff \simeq 1$.
In this narrow range
the effective reaction strength 
parameter is far away from either of its limiting values.
In any case, these deviations are insignificantly small.   

In the case of the dimerization-dissociation process the 
moment equations (\ref{eq:moment2}) 
are a set of three linear coupled differential equations.
As opposed to the case of 
dimerization without dissociation, here the equation for
$\ND$
does not only receive input from the other two equations, but
also generates an output into those equations.
This does not enable one to solve the first and third equations independently 
to obtain a time dependent solution as shown in the previous Section.
Here the time dependent solution will include three characteristic time scales
for the relaxation times of both
$\NA$, 
$\ND$ 
and 
$\R$.
To obtain these time scales we first write 
Eqs. (\ref{eq:moment2}) in
matrix form as

\begin{equation}
\dot{ \vec{ N}} = {\bf M} \vec N + \vec b,
\label{eq:moment2_matrixform}
\end{equation}

\noindent
where
$\vec N = (\NA, \ND, \R)$,
$\vec b = (g, 0, 0)$
and

\begin{eqnarray}
{\bf M} = 
\left(
\begin{array}{ccc}
- d_1    &  2 u        & -2
\\
0        & -(u + d_2)  &  1
\\
2 a g    & 2 a u       & -2(d_1 + a)
\end{array}
\right).
\label{eq:M2}
\end{eqnarray}

\noindent
The time dependent solution of the moment equations is given by

\begin{equation}
N_i = N_i^{\rm ss} + 
\sum_{j=1}^{3}
{{\bf C}_{ij} e^{-{t}/{\tau_j}}},
\label{eq:moment2_time_solution}
\end{equation}

\noindent
where 
$i,j = 1,2,3$.
Here, 
$\vec N^{\rm ss} = (\NAss, \NDss, \Rss)$,
and the matrix elements
${\bf C}_{ij}$
are determined by the initial conditions of the system.
The relaxation times 
$\tau_j$
are

\begin{equation}
\tau_j = - \frac{1}{\lambda_j},
\label{eq:tau_i}
\end{equation} 

\noindent
where
$\lambda_j$,
$j=1,2,3$,
are the eigenvalues of the matrix 
${\bf M}$ 
[Eq. (\ref{eq:M2})].
The time dependent solution obtained from the 
moment equations applies in the limits
where
$N_0 \ll 1$,
or in the limits where
$N_0 \gg 1$
and
$\gammaeff \ll 1$.
In the limit where
$N_0 \gg 1$
and
$\gammaeff \gg 1$,
the time dependent solution should be 
obtained from the rate equations
[Eqs. (\ref{eq:rate2})].

\section{Hetero-dimer Production}
\label{sec4}

Consider the case where the reacting monomers 
are from two different types of molecules,
$A$
and
$B$.
Each of these molecules is generated at a rate 
$g_A$ ($g_B$) 
and degraded at a rate 
$d_A$ ($d_B$).
The two molecules react to form the dimer
$D = AB$
at a rate
$a$ (s$^{-1}$).
The dimer product undergoes degradation at a rate 
$d_D$ (s$^{-1}$).
For simplicity, here we assume that the 
process of the dimer dissociation is suppressed.
The chemical processes in this system are thus

\begin{eqnarray}
&& \varnothing \arrow{g_A} A  \nonumber \\
&& \varnothing \arrow{g_B} B  \nonumber \\
&& A \arrow{d_A} \varnothing  \nonumber \\
&& B \arrow{d_B} \varnothing  \nonumber \\
&& A + B \arrow{a} D          \nonumber \\
&& D \arrow{d_D} \varnothing.
\label{eq:processes3}
\end{eqnarray}

\noindent
The average copy numbers of the reactive monomers and of the dimer product
are described by the following set of rate equations

\begin{eqnarray}
\frac{d\NA}{dt} &=& g_A - d_A \NA - \R  \nonumber \\
\frac{d\NB}{dt} &=& g_B - d_B \NA - \R  \nonumber \\
\frac{d\ND}{dt} &=& -d_D \ND + \R,
\label{eq:rate3}
\end{eqnarray}

\noindent
where 
$\R$, 
the dimer production rate, is given by

\begin{equation}
\R = a \NA \NB.
\label{eq:Rrate3}
\end{equation}

The master equation for this system describes the time evolution 
of the probabilities 
$P(N_A,N_B,N_D)$
for a population
$N_A$ 
molecules of type
$A$,
$N_B$ 
molecules of type
$B$
and
$N_D$ 
dimers
$D$
in the system.
It takes the form

\begin{eqnarray}
\label{eq:master3}
\frac{dP(N_A,N_B,N_D)}{dt} &=&
g_A [P(N_A - 1,N_B,N_D) - P(N_A,N_B,N_D)]
\nonumber \\ &+&
g_B [P(N_A,N_B - 1,N_D) - P(N_A,N_B,N_D)]
 \\ &+&
d_A [(N_A + 1)P(N_A + 1,N_B,N_D) - N_A P(N_A,N_B,N_D)]
\nonumber \\ &+&
d_B [(N_B + 1)P(N_A,N_B + 1,N_D) - N_B P(N_A,N_B,N_D)]
\nonumber \\ &+&
d_D [(N_D + 1)P(N_A,N_B,N_D + 1) - N_D P(N_A,N_B,N_D)]
\nonumber \\ &+&
a [(N_A + 1)(N_B + 1)P(N_A + 1,N_B + 1,N_D - 1) - N_A N_B P(N_A,N_B,N_D)]
\nonumber
\end{eqnarray}
 
\noindent
In the stochastic description the production rate of the dimer
$D$
is proportional to the number of pairs of 
$A$ and $B$
molecules in the system, namely

\begin{equation}
\R = a \NANB
\label{eq:Rmoment3}.
\end{equation}

\noindent
A more compact stochastic description can be 
obtained from the moment equations.
Here one must include equations for the first moments
$\NA$, $\NB$ and $\ND$,
and for the production rate, which involves the second moment
$\NANB$.
The results for the first moments can be obtained by tracing over the
master equation as shown in 
Sec. \ref{sec2}.
However, when deriving the equation for 
$\R$
one obtains

\begin{eqnarray}
\frac{d\R}{dt} 
&=& ag_B\NA + ag_A\NB - (d_A + d_B)\R   
\nonumber \\
&-& a^2(\NAsNB + \NANBs - \NANB),
\label{eq:Rmoment3_not_truncated}
\end{eqnarray}

\noindent
which includes third moments for which we have no equations.
To obtain the closure condition 
we follow the procedure presented in Sec. \ref{sec2}
and impose highly restrictive cutoffs on the master equation.
Here the cutoffs are chosen as
$N_A^{\rm max} = N_B^{\rm max} = N_D^{\rm max} = 1$.
These are the minimal cutoffs that enable the 
dimerization process to take place.
Under these cutoffs, 
the third order moments appearing in 
Eq. (\ref{eq:Rmoment3_not_truncated})
can be expressed by 
\cite{Barzel2007} 

\begin{equation}
\NAsNB = \NANBs = \NANB.
\label{eq:mixed_moments}
\end{equation}

\noindent
One then obtains a closed set of moment equations

\begin{eqnarray}
\frac{d\NA}{dt} &=& g_A - d_A\NA - \R 
\nonumber \\
\frac{d\NB}{dt} &=& g_B - d_B\NB - \R 
\nonumber \\
\frac{d\ND}{dt} &=& -d_D \ND + \R
\nonumber \\
\frac{d\R}{dt}  &=& ag_B\NA + ag_A\NB - (d_A + d_B + a)\R.
\label{eq:moment3}
\end{eqnarray}

\noindent
As in the case of the homo-molecular dimerization presented above, 
the validity of the moment equations 
extends well beyond the cutoff restriction.
It can be characterized by four parameters.
The first two are
$N_0^A = g_A/d_A$
and 
$N_0^B = g_B/d_B$,
which provide the upper limits 
on the monomer population sizes
$\NAss$
and
$\NBss$,
respectively.
The second two parameters are the reaction strength parameters,
which in the case of hetero-dimer production are
$\gamma_A = a g_A /(d_A d_B)$
and
$\gamma_B = a g_B /(d_A d_B)$.
In the limit where the populations are 
small, the moment equations provide accurate 
results for all the moments appearing in 
Eqs. (\ref{eq:moment3}).
When the populations are large, 
the moment equations provide accurate results 
for the dimerization rate,
$\R$, 
and for the dimer population
$\ND$.
However, if the reaction strength parameters are also large,
the moment equations will not correctly account 
for the monomer population sizes,
$\NA$
and 
$\NB$.

In Fig. \ref{fig9} we present 
$\NAss$ (circles),
$\NBss$ (squares),
$\NDss$ (triangles)
and
$\Rss$ ($\times$),
versus the reaction strength parameters
$\gamma_A = \gamma_B$,
as obtained from the moment equations.
Here the parameters are
$g_A = 10^{-2}$,
$g_B = 10^{-2}$,
$d_A = 1$,  
$d_B = 10$,
$d_D = 0.2$ 
(s$^{-1}$),
and the parameter
$a$
is varied.
These parameters are within
the limit of small populations.
The results are in perfect agreement 
with those obtained from the master 
equation (solid lines).
The rate equations (dashed lines) 
show strong deviations, which are mainly 
expressed in the reaction-dominated regime. 
In Fig. \ref{fig10} we present 
$\NAss$ (circles),
$\NBss$ (squares),
$\NDss$ (triangles)
and
$\Rss$ ($\times$),
versus the reaction strength parameters
$\gamma_A = \gamma_B$,
as obtained from the moment equations.
Here the parameters are
$g_A = 10^{3}$,
$g_B = 10^{3}$,
$d_A = 1$,  
$d_B = 10$, 
$d_D = 0.2$ 
(s$^{-1}$),
and the parameter
$a$
is varied.
These parameters are within the limit of large populations.
Nevertheless the results obtained from the moment equations 
for 
$\NDss$
and for
$\Rss$
are in good agreement 
with those obtained from the rate equations (dashed lines)
in both the reaction-dominated limit 
and in the degradation-dominated limit.
For the monomer population sizes,
$\NAss$
and 
$\NBss$,
the moment equations apply only in the limit where
$\gamma_{A} < 1$
and
$\gamma_{B} < 1$.

\section{Summary and Discussion}
\label{sec5}

We have addressed the problem 
of dimerization reactions
under conditions in which fluctuations are important.
We focused on two types of reactions, 
homo-molecular dimerization ($A+A \rightarrow A_2$) 
and hetero-dimer
production
($A+B \rightarrow AB$).
Common approaches for the stochastic simulation of such 
reaction systems include the direct integration of the 
master equation and Monte Carlo simulations.
The master equation involves a large number of coupled equations,
for which there is no analytical solution in the time-dependent case.
Monte Carlo simulations are often computationally intensive and 
require averaging over large sets of data.
As a result,
the relaxation times 
and the steady state populations
for given values of the rate constants
can only be obtained by numerical calculations.

Here we have utilized the recently proposed moment equations method,
in order to obtain an analytical solution for 
the relaxation times and for the steady state populations.
The moment equations provide an accurate description of 
dimerization processes in the stochastic limit, at the cost
of no more than three or four coupled linear differential
equations.
Another useful feature of these 
equations is that in certain cases they also apply
in the deterministic limit.
Using the moment equations we 
obtained a complete time dependent solution for
the monomer population 
$\NA$, 
the dimer population 
$\ND$ 
and the dimerization rate 
$\R$,
in the case of homo-molecular dimerization.
Expressions for the relaxation times 
and the steady state populations were
found in terms of the rate constants of the 
different processes.
In the case of hetero-dimer production the moment 
equations include four coupled linear
equations.
These equations can be easily solved by direct numerical
integration.
However, a general algebraic expression 
for this solution is tedious 
and was not pursued in this paper.
Stochastic dimerization processes appear in many natural systems.
Below we discuss several examples.

One of the most fundamental chemical 
reactions taking place in the interstellar medium is
hydrogen recombination, namely 
${\rm H} + {\rm H} \rightarrow {\rm H}_2$
\cite{Gould1963,Hollenbach1970,Hollenbach1971a,Hollenbach1971b}.
This reaction occurs on the surfaces of microscopic dust grains
in interstellar clouds
\cite{Spitzer1978,Hartquist1995,Herbst1995}.
The resulting 
${\rm H}_2$
molecules 
participate in further reactions
in the gas phase, 
giving rise to more complex molecules
\cite{Tielens2005}.
They also play an important role in 
cooling processes during gravitational collapse and star formation. 
In recent years there has been much activity in the computational
modeling of interstellar chemistry.
While the gas phase chemistry can be simulated by rate equations
\cite{Pickles1977,Hasegawa1992},
the reactions taking place on the dust grain 
surfaces often require stochastic methods
\cite{Biham2001,Green2001,Charnley2001}.
This is because under the extreme interstellar 
conditions of low gas density,
the population sizes of the reacting H atoms on 
the surfaces of these microscopic grains
are small and highly fluctuative
\cite{Tielens1982,Charnley1997,Caselli1998,Shalabiea1998}.
The processes taking place on the grains are
the accretion of H atoms onto the surface,
the desorption of H atoms from the surface, 
and the diffusion of atoms between adsorption 
sites on the surface.
These processes can be described by the 
dimerization system discussed in Sec. \ref{sec2}.
In recent years, experimental work was 
carried out in an effort to obtain the relevant 
rate constants 
and for certain grain 
compositions these constants were found
\cite{Pirronello1997a,Katz1999,Hornekaer2003,Perets2005,Perets2007}.
The solution of the moment equations, 
as appears in Sec. \ref{sec2},
provides the production rate of molecular hydrogen
on interstellar dust grains,
in the limit of small grains and low fluxes,
where fluctuations are important.

In the biological context, 
regulation processes in cells can be described by
networks of interacting genes
\cite{Alon2006,Palsson2006}.
The interactions between genes  
include transcriptional regulation processes as well as protein-protein
interactions
\cite{Yeger-Lotem2004}.
Due to the small size of the cells, 
some of these proteins my appear
in low copy numbers, with large fluctuations
\cite{McAdams1997,Paulsson2000,Paulsson2004,Friedman2006}. 
Deterministic methods are thus not suitable for the
modeling of these systems.
Dimerization of proteins is a common process
in living cells. 
In particular, many of the transcriptional regulator
proteins bind to their specific promoter sites on the DNA
in the form of dimers.
It turns out that such dimerization, taking place before
binding to the DNA, provides an effective mechanism for
the reduction of fluctuations in the monomer copy numbers
\cite{Bundschuh2003}.

In a broader perspective, complex reaction 
networks appear in a variety of physical contexts.
The building blocks of these networks are intra-species 
interactions and inter-species interactions.
Thus, the analysis presented in this paper of 
homo-molecular and hetero-molecular dimerization 
processes, lays the foundations for the analysis of more complex networks.
Complex stochastic networks are difficult to simulate
using standard methods,
because they require exceedingly long simulation times.
The moment equations,
applied here to dimerization systems, 
provide a highly efficient method for the
simulation of complex chemical networks.

This work was supported by the US-Israel Binational
Science Foundation and by the France-Israel High
Council for Science and Technology Research.


\clearpage
\newpage


\begin{table}
\caption{
The validity
($\surd$) or invalidity ($\times$) 
of the moment equations for the
evaluation of 
$\NA$,
$\ND$
and 
$\R$
in the different limits
of the dimerization system.
The results for
$\ND$
and 
$\R$
are valid in all limits.
The results for
$\NA$ are invalid 
in the limit of large systems and reaction-dominated
kinetics.
}
\begin{tabular}{||c||c|c||c|c||}
\hline
\hline
 & \multicolumn{2}{c||}{$N_0 \ll 1$} & \multicolumn{2}{c||}{$N_0 \gg 1$} 
\\
\cline{2-5}
 & $\gamma \ll 1$ & $\gamma \gg 1$ & $\gamma \ll 1$ & $\gamma \gg 1$
\\
\hline
\hline
$\NA$ & $\surd$ & $\surd$ & $\surd$ & $\times$ 
\\
\hline
$\ND$ & $\surd$ & $\surd$ & $\surd$ & $\surd$ 
\\
\hline
$\R$  & $\surd$ & $\surd$ & $\surd$ & $\surd$ 
\\
\hline \hline
\end{tabular}
\label{tab1}
\end{table}


\begin{table}
\caption{
The relaxation times for 
$\NA$
and 
$\ND$
in the different limits of the dimerization system.
For the monomer population the 
relaxation time is determined by
the degradation rate,
$d_1$ in three of the four limits.
In the reaction-dominated, large system limit 
the relaxation time is determined by
$\sqrt{8ag}$.
In case that the degradation rate of the dimer, $d_2$,
is large, its relaxation follows that of the monomer
population. Otherwise, it is determined by $d_2$.
}
\begin{tabular}{||c||c|c||c|c||}
\hline
\hline
& \multicolumn{2}{c||}{$N_0 \ll 1$} & \multicolumn{2}{c||}{$N_0 \gg 1$} 
\\
\cline{2-5}
& $\gamma \ll 1$ & $\gamma \gg 1$ & $\gamma \ll 1$ & $\gamma \gg 1$
\\
\hline
\hline
$\tau_A$  
& \multicolumn{3}{c|}{$\frac{1}{d_1}$}
& $\frac{1}{\sqrt{8ag}}$ 
\\
\hline
$\tau_D$ 
& \multicolumn{3}{c|}{$\max \left(\frac{1}{d_1},\frac{1}{d_2}\right)$}
& $\max \left(\frac{1}{\sqrt{8ag}},\frac{1}{d_2}\right)$ 
\\
\hline \hline
\end{tabular}
\label{tab2}
\end{table}


\begin{figure}
\includegraphics[width=12cm]{fig4.eps}
\caption{
(color online)
The average monomer copy number,
$\NAss$ (circles),
the average dimer copy number,
$\NDss$ (squares)
and the dimerization rate,
$\Rss$ (triangles)
versus the reaction strength parameter,
$\gamma$,
as obtained from the moment equations
under steady state conditions.
The parameters used here satisfy the large
system limit,
namely
$N_0 = 10^{4}$.
In this limit the rate equations
(dashed lines) are reliable.
The moment equation results are in agreement 
with the results obtained
from the rate equations for the dimer copy number 
and its production rate.
This is in spite of the fact that the copy numbers are
large, far beyond the conditions for which the moment equations
are designed.
For the monomer copy number, 
the moment equations deviate in the
reaction-dominated limit 
($\gamma > 1$).
Slight deviations in the results for 
$\NDss$ 
and 
$\Rss$
appear around 
$\gamma \simeq 1$,
where the crossover from the degradation-dominated regime 
to the reaction-dominated regime takes place.
}
\label{fig4}
\end{figure}


\begin{figure}
\includegraphics[width=12cm]{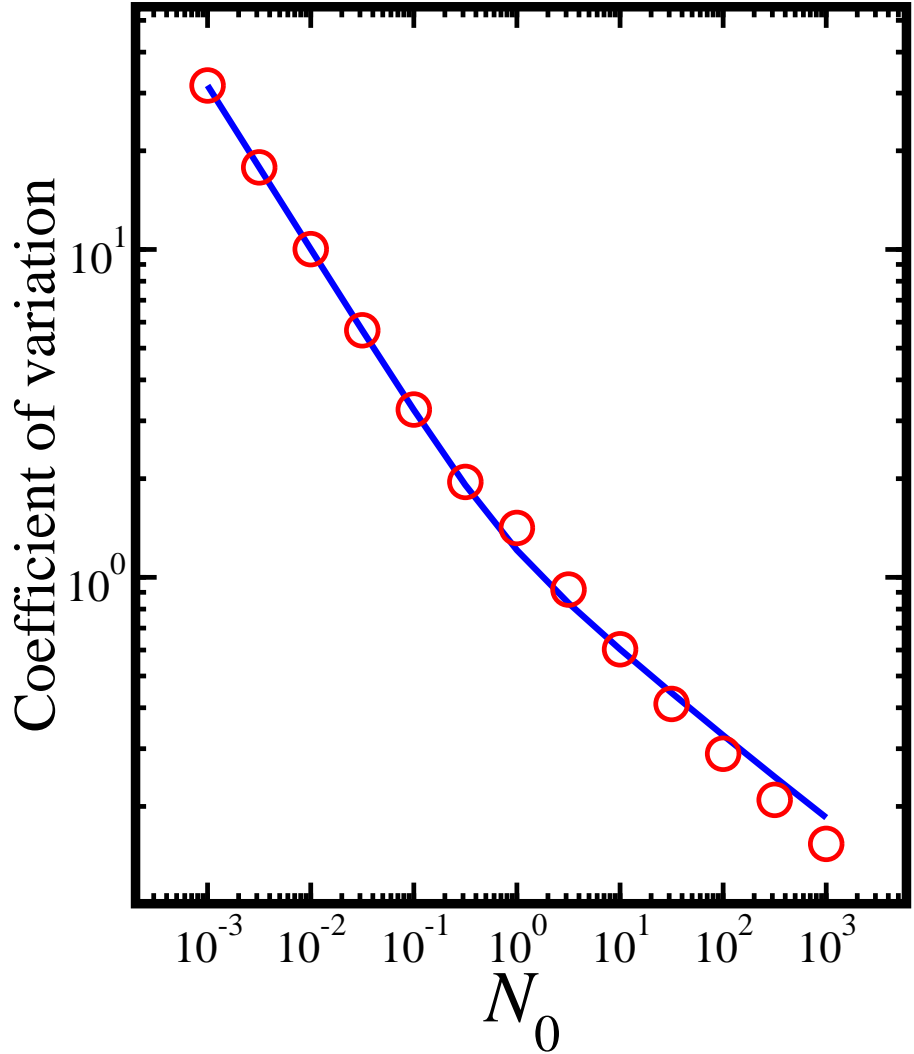}
\caption{
(color online)
The coefficient of variation of 
the monomer copy number,
$\sigmass / \NAss$, 
versus the system size parameter
$N_0$ (circles),
as obtained from 
Eq. (\ref{eq:moment_rate_sigma}). 
As the average population size increases,
$\sigmass$
becomes smaller than
$\NAss$.
For 
$N_0 < 1$
the results are in perfect agreement 
with those obtained from the 
master equation (solid line).
For 
$N_0 > 1$,
where the expression in 
Eq. (\ref{eq:moment_rate_sigma}) 
combines results obtained from 
the moment equations and the rate equations, 
a slight deviation emerges.
Nevertheless, Eq. (\ref{eq:moment_rate_sigma}) 
is shown to provide a good 
approximation for 
$\sigmass$.
}
\label{fig5}
\end{figure}


\begin{figure}
\includegraphics[width=12cm]{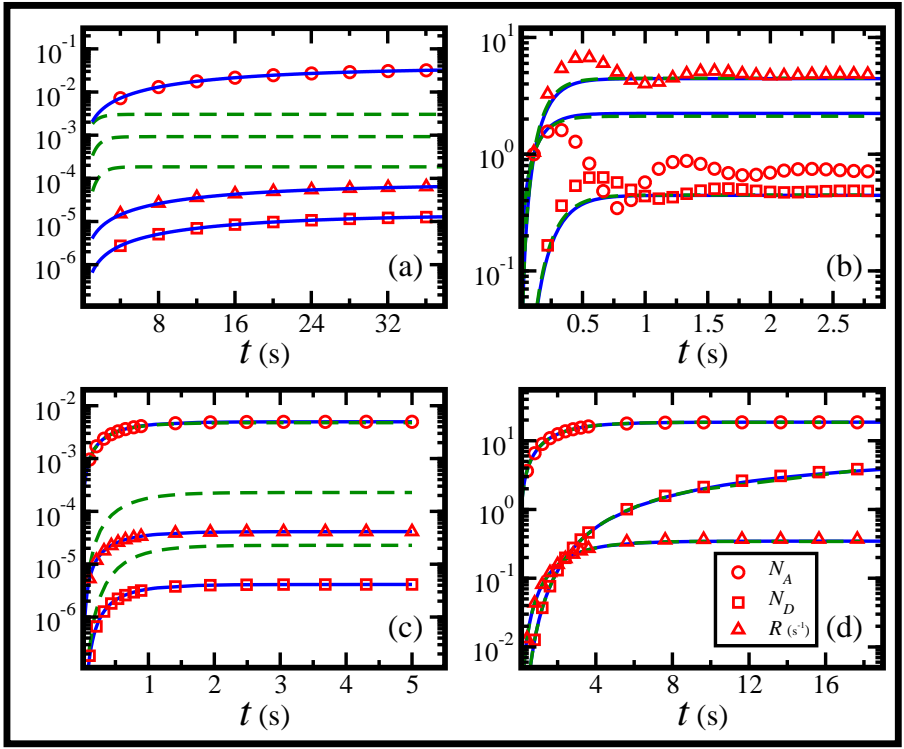}
\caption{
(color online)
The time dependence of 
$\NA$ (circles),
$\ND$ (squares)
and 
$\R$ (triangles),
as obtained from the moment equations.
Four different limits are observed.
In the limit where 
$N_0 \ll 1$
and
$\gamma \gg 1$ (a)
the moment equations are in perfect agreement with the master 
equation.
In this limit, the rate equations fail to account both for the
steady state copy numbers and for the relaxation time.
According to the rate equations, 
the relaxation time is
$\tau \simeq 1/\sqrt{8ag}$,
while according to the moment equations it is 
$\tau \simeq 1/d_1$,
in agreement with the master equation.
In the limit where 
$N_0 \gg 1$
and
$\gamma \gg 1$ (b),
the rate equations and the master equation are in good agreement.
In this limit the moment equations fail to 
produce the correct time dependent 
solution, as they predict an oscillatory convergence towards steady state.
Note, however, that even in this limit, 
the moment equations do provide correct results for 
the steady state values of
$\ND$
and 
$\R$.
In the limit where 
$N_0 \ll 1$
and
$\gamma \ll 1$ (c),
the moment equations are in perfect agreement with the master 
equation (solid lines).
The rate equations (dashed lines),
which do not account for fluctuations, 
fail in this limit.
The limit where 
$N_0 \gg 1$
and
$\gamma \ll 1$ 
is shown in (d).
Although the copy numbers are large in this limit,
the moment equations are still in perfect agreement 
with the master equation.
Not surprisingly, the rate equations also apply.
}
\label{fig6}
\end{figure}


\begin{figure}
\includegraphics[width=12cm]{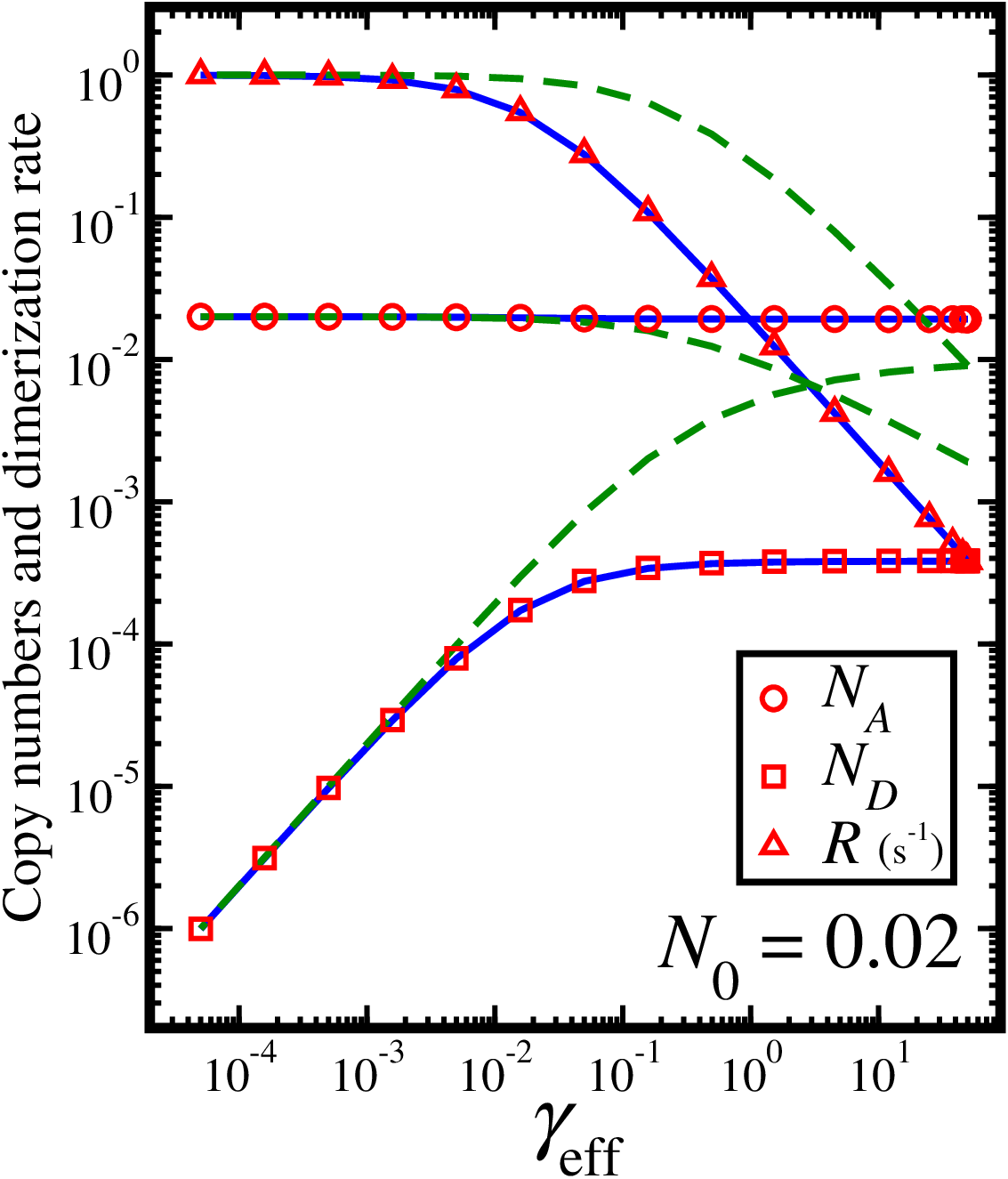}
\caption{
(color online)
The average monomer copy number,
$\NAss$ (circles),
the average dimer copy number,
$\NDss$ (squares)
and the dimerization rate,
$\Rss$ (triangles),
versus the effective reaction strength parameter,
$\gammaeff$,
as obtained from the moment equations
for the dimerization-dissociation reaction.
The parameters used here represent the small system limit,
namely
$N_0 = 0.02$.
The moment equation results are in perfect 
agreement with those obtained
from the master equation (solid lines).
However the results of the rate equations (dashed lines)
show significant deviations.
}
\label{fig7}
\end{figure}


\begin{figure}
\includegraphics[width=12cm]{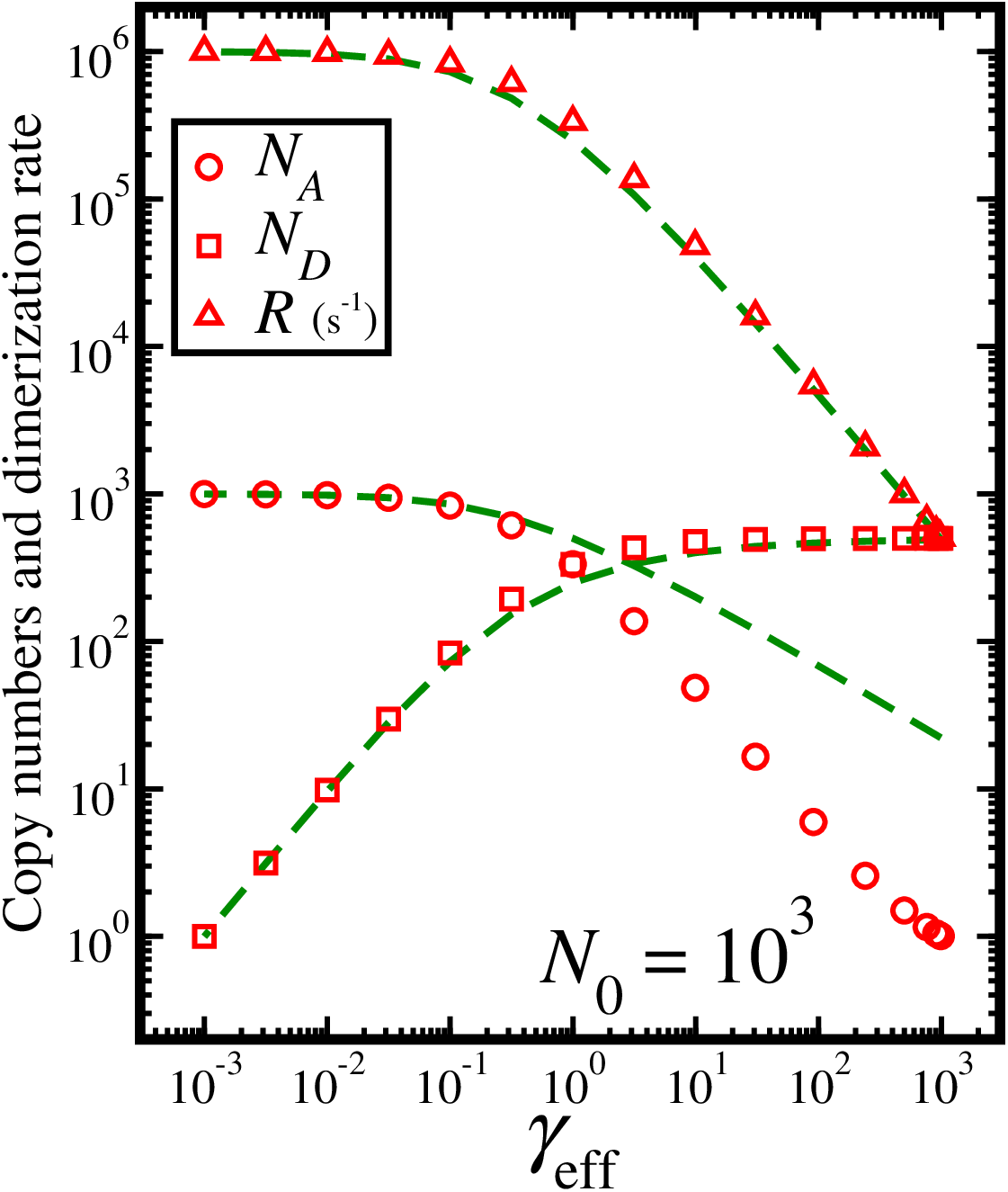}
\caption{
(color online)
The average monomer copy number,
$\NAss$ (circles),
the average dimer copy number,
$\NDss$ (squares)
and the dimerization rate,
$\Rss$ (triangles),
versus the effective reaction strength parameter,
$\gammaeff$,
as obtained from the moment equations
for the dimerization-dissociation reaction.
The parameters used here represent the large system limit,
namely
$N_0 = 10^{3}$.
In this limit the rate equations
(dashed lines) are reliable.
The moment equation results are in agreement with those obtained
from the rate equations for the dimer copy number and its production rate.
However, for the monomer copy number, 
the moment equations deviate in the
reaction-dominated limit 
($\gammaeff > 1$).
Slight deviations in the results for 
$\NDss$ 
and 
$\Rss$
appear around 
$\gammaeff \simeq 1$,
where the crossover from the degradation-dominated regime to 
the reaction-dominated regime takes place.
}
\label{fig8}
\end{figure}


\begin{figure}
\includegraphics[width=12cm]{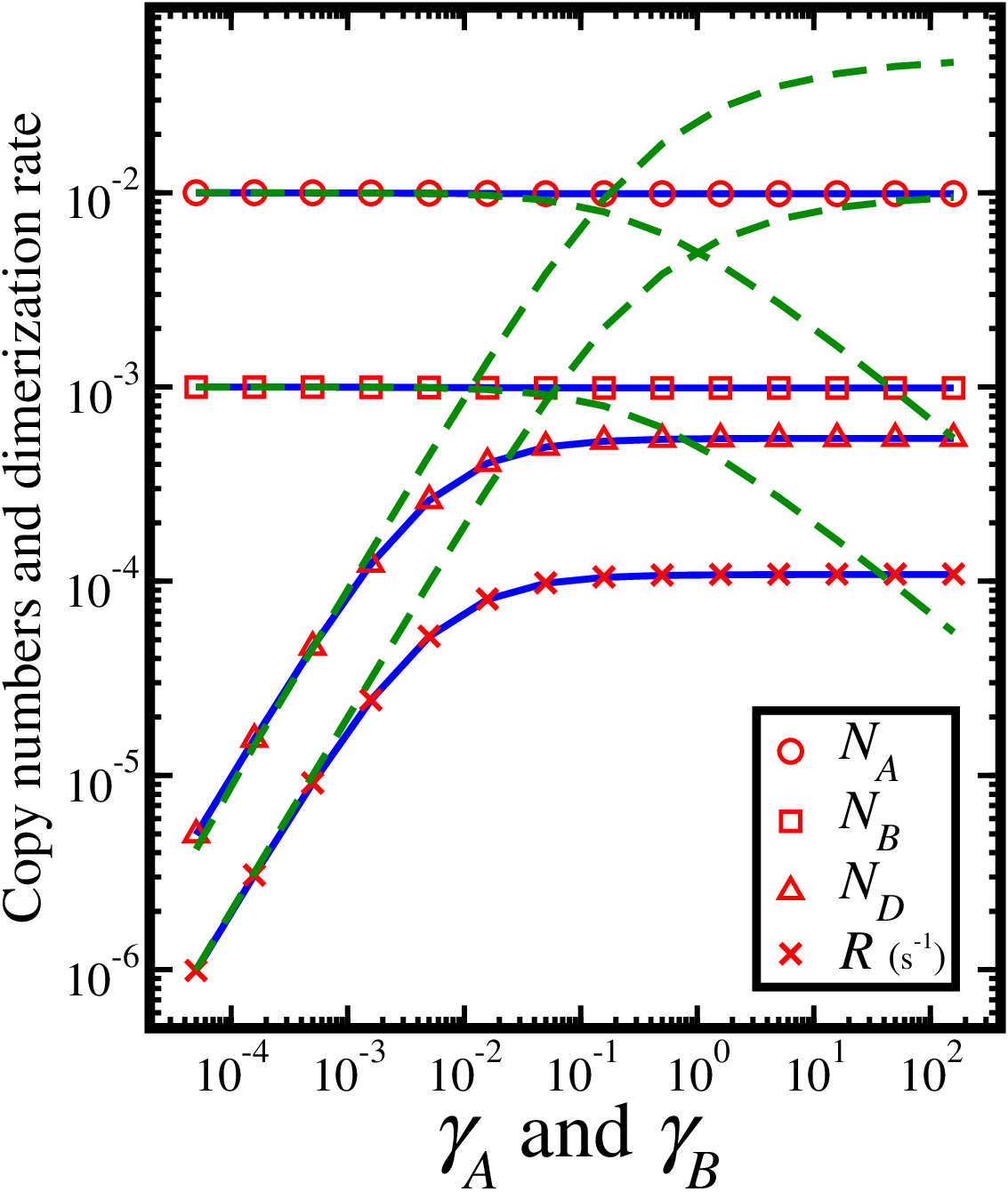}
\caption{
(color online)
The average monomer copy numbers,
$\NAss$ (circles)
and 
$\NBss$ (squares),
the average dimer copy number,
$\NDss$ (triangles),
and the dimerization rate,
$\Rss$ ($\times$),
versus the reaction strength parameters,
$\gamma_A = \gamma_B$,
as obtained from the moment equations,
for the hetero-dimer production system.
The parameters used here represent the small system limit.
The moment equation results are in perfect agreement with those obtained
from the master equation (solid lines).
However the results of the rate equations (dashed lines)
show significant deviations.
}
\label{fig9}
\end{figure}


\begin{figure}
\includegraphics[width=12cm]{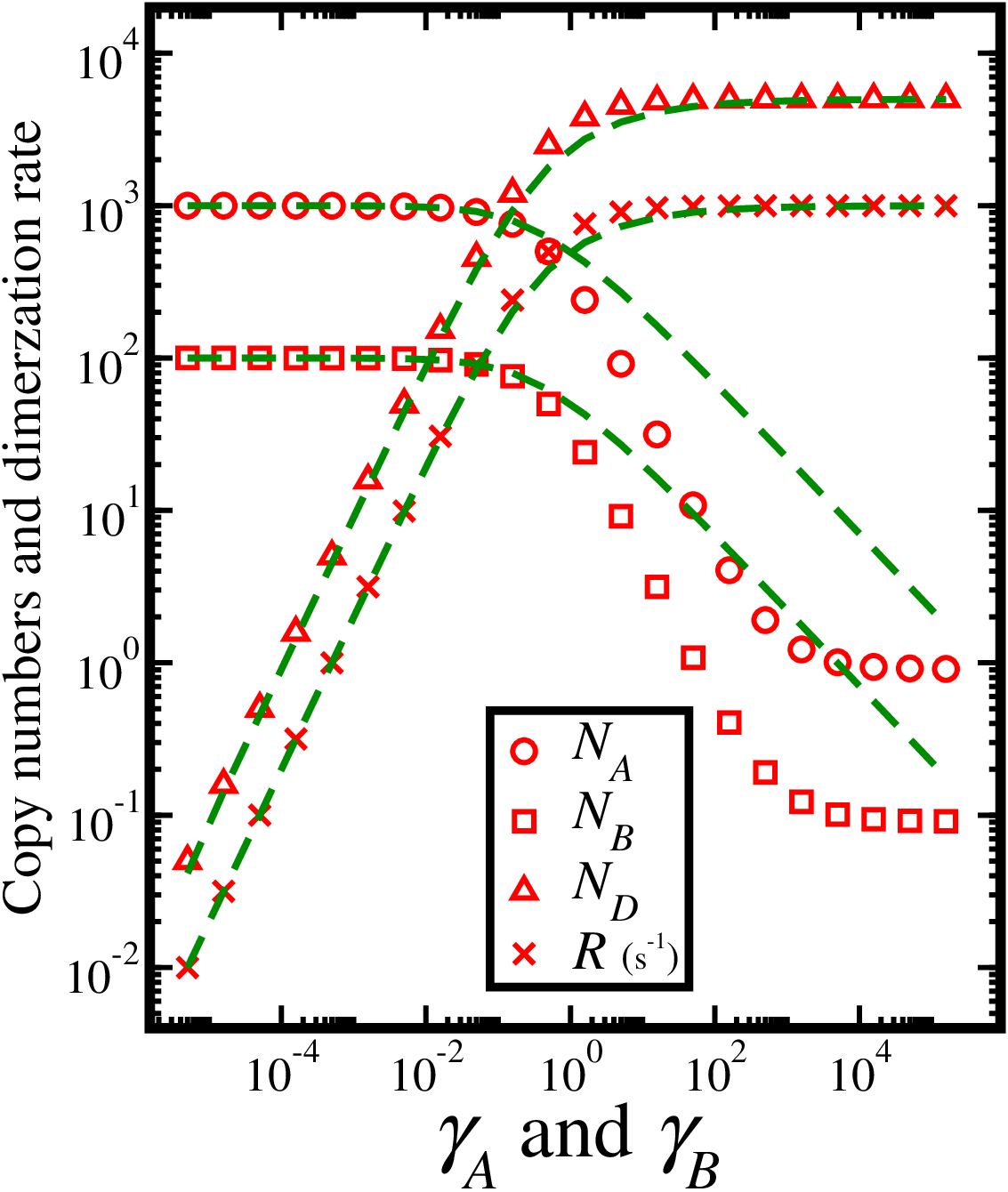}
\caption{
(color online)
The average monomer copy numbers,
$\NAss$ (circles)
and 
$\NBss$ (squares),
the average dimer copy number,
$\NDss$ (triangles),
and the dimerization rate,
$\Rss$ ($\times$),
versus the reaction strength parameters,
$\gamma_A = \gamma_B$,
as obtained from the moment equations
for the hetero-dimer production system.
The parameters used here represent the large system limit.
In this limit the rate equations
(dashed lines) are reliable.
The moment equation results are in agreement with the results obtained
from the rate equations for the dimer population and its production rate.
However, for the monomer copy numbers, 
the moment equations deviate in the
reaction-dominated limit 
($\gamma_{A(B)} > 1$).
}
\label{fig10}
\end{figure}

\end{document}